\documentclass[10pt,conference]{IEEEtran}

\usepackage{algorithm}
\usepackage{algpseudocode}
\usepackage{float}
\usepackage{xcolor}
\usepackage{hyperref}
\usepackage{graphicx}
\usepackage{listings}
\usepackage[inline]{enumitem}

\usepackage{comment}

\begin{document}

\title{From Benchmark Data To Applicable Program Repair: An Experience Report}

\author{
\IEEEauthorblockN{Mahinthan Chandramohan, Jovan Jancic, Yuntong
Zhang\textsuperscript{1}\thanks{\textsuperscript{1}Currently at the National
University of Singapore.} and
Padmanabhan Krishnan}
   \IEEEauthorblockA{Oracle Labs\\
        Brisbane, Australia\\
\{mahin.chandramohan, jovan.jancic, paddy.krishnan\}@oracle.com}
}

\IEEEoverridecommandlockouts

\maketitle

\begin{abstract}
This paper describes our approach to automated program repair. We combine
various techniques from the literature to achieve this. Our experiments show
that our approach performs better than other techniques on standard benchmarks.
However, on closer inspection, none of these techniques work on realistic
defects that we see in industry.

We find that augmenting code with formal specifications enables LLMs to generate
higher-quality unit tests, especially for complex production code
with improved coverage of edge cases and exception handling. However, specifications add little value for well-understood errors (e.g., null pointer, index out of bounds), but are beneficial for logic and string manipulation errors. Despite encouraging benchmark results, real-world adoption is limited since passing tests do not guarantee correct patches.
Current challenges include insufficient expressiveness of the JML specification
language, necessitating advanced verification tools and richer predicates.
Our ongoing work is exploring contract automata, programming by example, and test-case repair, with a focus on integrating human feedback and measuring productivity gains — highlighting the gap between academic benchmarks and practical industry needs.
\end{abstract}


\section{Introduction}

Automated program repair has been studied extensively. The initial techniques
to repair defects
used various ideas from software engineering and program analysis.
These include defect localisation by perhaps using failing tests and other
information such as changes to the code which could have introduced the defect,
classifying the type of defect to see what types of repairs could be
feasible (e.g., template based repair, constraint-based repair),
actual patch generation
and test generation to validate the generated patch.
Overall, the aim of program repair is to search the space of program fragments
and identify the program fragments that can be part of the patch
\cite{LPR-cacm19}.

More recently, machine learning techniques using large language models (LLMs)
have been used in the process \cite{chatrepair,thinkrepair,AutoCodeRover}.
Given the complexity of finding a repair once a
defect has been reported, LLM-based agent systems are also becoming popular
\cite{RepairAgent24,AgentProgramRepair25}.
For instance, RepairAgent \cite{RepairAgent24} interacts with a set of tools to
analyse and fix bugs. While the LLM decides how to use the tools, a state
machine is used to provide guidance to the LLM. Also, the tools they describe
are relatively simple (e.g., searching the code for relevant classes or
methods). In practice,
for this to be effective, one has to combine ideas from program analysis
with LLMs \cite{InferFix23,LogicPy}.

In this paper we present our solution that can be used with
agent-based systems that addresses
the different aspects of program repair. In the next section we outline the
various steps that can be involved in automated program repair which motivates
our solution which is described in Section~\ref{sec:ourContrib}. This solution
includes bug localisation, test generation, use of suitable contexts and
specification generation.

A key contribution of our work is described in 
Section~\ref{sec:eval} where we examine the effectiveness of our approach on
standard benchmark programs and how applicable are these techniques to real
situations. We identify shortcomings in using aggregrated results and
illustrate, via a few examples, the importance of the human in the loop. Thus
for real systems, the current approaches (including our own)
do not achieve automated program
repair; rather it is a tool that can potentially help improve productivity.

\section{General Approach}\label{sec:genApproach}

We now describe the typical process taken by a developer to fix a reported
defect.

\begin{enumerate*}[label=(\arabic*)]
\item
When a defect is reported, one must identify the root cause of the defect.
This is because the location where the defect
manifests (i.e., a crash) need not be
the root cause.
\item Furthermore, the reporter of the defect may not provide full
information, say because of sensitivity. Hence one needs to be able to reproduce
the defect.
\item Once the location of the defect has been identified, one needs to
needs to identify what is exactly incorrect. In manual approaches, the
developers rely on their insights into the code. However, if this
has to be automated some formality is required, e.g., information from
static analysers such as path conditions or explicit formal
specifications, if available.
\item 
Such constraints from the existing program will
characterise what is wrong, but they by themselves do not address what is
a correct specification. Hence techniques to generate a correct specification
need to identified.
\item Using the information from the previous steps, one has to generate the
actual patch.
\item Once the patch has been generated, we have to validate the patch. This
could use the tests that exposed the defect. 
However, we must also ensure that the generated patch neither introduces new defects nor overfits the defect report, so that it passes existing test cases and generalizes beyond them~\cite{le2018overfitting}
\end{enumerate*}

Although we present the process as a sequence, in practice, there are multiple
workflows that are possible. This is especially true when LLMs are used where
incorrect information from previous phases may need to be corrected. For this
one can use agents \cite{RepairAgent24,AgentProgramRepair25}.

Other aspects not directly considered in the general process
include trusting the code that is
generated \cite{TrustAbhik25}, validating the prompts \cite{PromptPex25} and
validating the agents themselves \cite{AgentMalicious25}. These are important
considerations before we can deploy our solutions in production.

\section{Our Contribution}\label{sec:ourContrib}

Our overall approach
aligns with the steps outlined in Section~\ref{sec:genApproach}. However,
there are a few key differences which has led to the development of specific
tools.
For instance, we could not use monolithic tools (e.g., static analysers)
directly. While they report defects along with associated abstract traces,
we needed access
to many of the internal details. Without the internal details, the LLMs are not
effective as they are not mature enough to handle coarse summarised information.
LLMs are, in general, incompatible with the tools designed for large-scale code, which were intended for human use without consideration for LLM integration.

We explain this via a specific example of defects that depend on
taint analysis. While, our static analysis tool reported that a
particular value was tainted, the reasons why they are tainted was not
available, except in the abstract trace.
But this is required for program repair as the sanitisation has to be
applied to the source of the taint. Listing~\ref{lst:psuedo_code} illustrates this. 

\begin{lstlisting}[caption={Taint propagation psuedo-code}, label={lst:psuedo_code}]
   String userInput;
   String getInput() {
      userInput = read();
   }
   
   void executeStatement() {
     y = "SELECT a from A WHERE" +
         userInput;
     jdbcCall(y);
   }
\end{lstlisting}

Static analysers will report that variable \lstinline|y| is tainted. But from a
repair perspective, we need to sanitise the field
\lstinline|userInput| to prevent a potential
SQL injection attack. For this we need both interprocedural and field-sensitive
analysis. It is the results of such analyses that LLMs are unable to infer
directly.
Therefore, in general, the internal details of various tools need to be
exposed.

While languages like CodeQL \cite{COdeQLIncr23}
can be used to write specifications with limited scope (e.g., field-sensitive
flow analysis) and then used in an incremental fashion,
licensing issues prevented from using it in our
explorations.

\begin{figure}[htb]
\centerline{\includegraphics[width=0.5\textwidth]{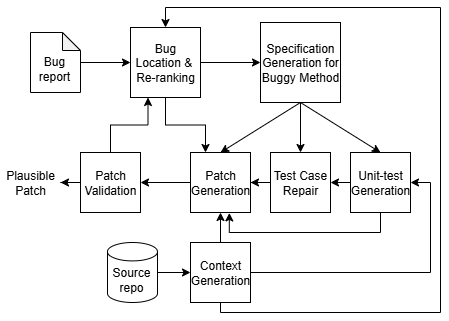}}
\caption{Proposed Automated Program Repair Workflow}
\label{fig:workflow}
\end{figure}

Based on the above observations, we use the
workflow depicted in Figure~\ref{fig:workflow}. We have developed
tools with a focused scope that address each step in the workflow.
These are described in the next sections.

\subsection{Bug Localisation and Reproduction}\label{sec:buglocalization}

We have developed a technique that identifies plausible bug locations
at commit, file and line level \cite{CNK-buglocalization24}. The
commit and line level information
enables the downstream tasks to work better than  purely
file-level localisation techniques \cite{chakraborty2023rlocator}.

Our approach uses a pre-trained model based on UniXCoder
\cite{guo2022unixcoder}
and uses LLMs to rank the proposed locations.
Our pre-trained model was developed before LLMs were accessible.
Pragmatics means that we cannot throw away what has been developed and we cannot
redo that work using only LLMs.
Also, as part of reproducing the bug, we use various test generation techniques
\cite{Issue2Test25} including unit test generation \cite{chen2024chatunitest}.

\subsection{Context Generation}\label{subsec:context}

For all the steps outlined earlier, it is important to give the LLMs the right
context \cite{liu2024codexgraph,qu2024context}. In our experience, the context
needs to have the third party public APIs that are invoked by the code. Hence
apart from the code in the application's repository, we also extract
the public
API of all the imported libraries 
via the 
Tree-sitter library\footnote{\url{https://tree-sitter.github.io/tree-sitter}}.
Other context information is specific to the task that is being performed.
Performance, at least the initial set up, is an important consideration. Since
context generation is similar to IDEs creating suitable indexes, we compared our
work with the community version of
IntelliJ\footnote{\url{https://www.jetbrains.com/idea/}}. The results for the
initial indexing/context creation are
presented in Table~\ref{tab:contextPerf}. The context is continuously updated via
monitoring the changes to the repository. As the time taken for
this is also comparable to, or even better than, the time taken by IntelliJ to update the
index, the
performance has been deemed to be acceptable by production groups within Oracle.

\begin{table}[htb]
\caption{Context Generation vs IntelliJ Indexing Time}
\label{tab:contextPerf}
\begin{tabular}{|p{0.15\textwidth}|p{0.07\textwidth}|p{0.06\textwidth}|p{0.07\textwidth}|}
\hline
\hline
Project Name & \# Methods & IntelliJ Indexing & Context
Generation \\
\hline
Spring-framework & 90,296 & 214.61s & 179.92s \\
Spring-boot & 49,255 & 197.96s & 99.11s \\
Micronaut-core & 21,145 & 149.81s & 43.64s\\
\hline
\end{tabular}
\end{table}

\subsection{Formal Specification Generation}\label{sec:formalSpecGen}

Adopting a specification-based technique requires tools that can work
on incomplete code. Hence we are unable most off-the-shelf static analysis
tools as they require the program to be compilable.
Our approach to specification generation,
which is similar to the techniques described
in \cite{ma2024specgen,FormalBench25},
includes both the failing and passing
tests, buggy method, and the list of callee methods reachable via the buggy
method in the context.

Since specifications are not available, we rely on the
LLM to generate them. However, we verify the generated specification 
using tools like OpenJML \cite{openjml}.

Our specification generation workflow assumes the following input: a buggy codebase, one or more failing tests, the buggy method name, as well as a configurable maximum iteration number.
We assume the buggy method to be fixed is known from the bug
localisation technique explained in Section~\ref{sec:buglocalization}.
It first retrieves the names of the methods that are called by the buggy method, where it invokes context generation (Section~\ref{subsec:context}) to obtain callee methods to the buggy method which are also defined in the target project codebase.
The callee methods together with the buggy method form the context for generating the initial specification.
Here, JML specifications for the callee methods are also generated so that the
verifier can be run on the buggy method later. 

After the initial JML specification has been generated by an LLM, our technique 
enters a \textit{specification refinement} loop. 
Here, the current JML specification is first verified by running it against the buggy codebase using the OpenJML~\cite{openjml} verification engine.
At this step, the verification can fail either because (1) the JML
specification has either syntactic or semantic errors, or (2) the JML specification correctly captures the intended behaviour of the buggy code so the verification engine reports errors.
In the refinement loop, we instruct the LLM to examine the verification output and improve the previous JML specification with respect to the first type of error (i.e., the JML specification has syntactic or semantic errors).

The procedure terminates and returns the final JML specification when either no further syntactic or semantic errors are detected and OpenJML successfully verifies the specification, or resource limits are reached.

\begin{figure*}[t]
    \includegraphics[width=\textwidth]{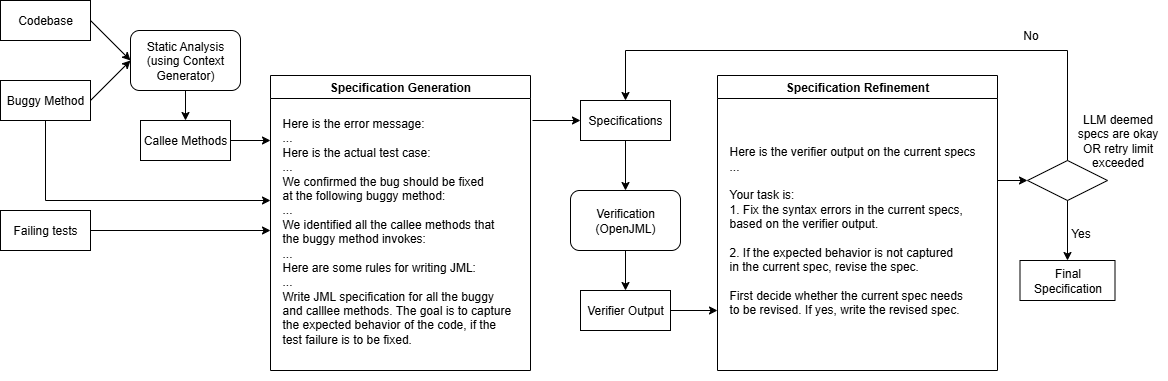}
    \caption{Formal Specification Generation and Refinement Workflow}
    \label{fig:spec_gen_workflow}
\end{figure*}

Figure~\ref{fig:spec_gen_workflow} illustrates the workflow of the formal specification generation and refinement process. It also includes the example prompts (simplified) for specification generation and refinement.

\subsection{Unit Test Generation}

The formal specification described earlier is useful in our approach to
generate unit tests. That is, by augmenting the code with the specification, the
LLM is able to generate better unit tests. These include test cases that
exercised edge cases and generated exceptions (which is often the case for buggy
behaviour). Based on an experiment with the Oracle Code Assist
team\footnote{\url{https://www.oracle.com/au/application-development/code-assist/}}
on cases which were deemed to be challenging, our tool
was able to generate better unit
tests for about 68\% of methods as opposed to around 42\% of methods that the
various techniques were able to do so. The quality of the tests were evaluated
manually by the owners of the code-base.

\subsection{Patch Generation}

We extend CigaR~\cite{Cigar} with specification generation. In particular, we use the bug description, the buggy method, and the generated JML specifications to produce plausible patches. That is, in addition to instructing the LLM to generate candidate patches, we include the auto-generated JML specification when constructing the prompt for patch generation.

In this workflow, we first attempt to generate the patch using the original
prompt without specifications (plain mode). If this attempt is unsuccessful, we
then extend the prompt with JML specifications (mixed mode) that has been
generated. 

\begin{figure}[t]
   \centering
   \includegraphics[width=0.51\textwidth]{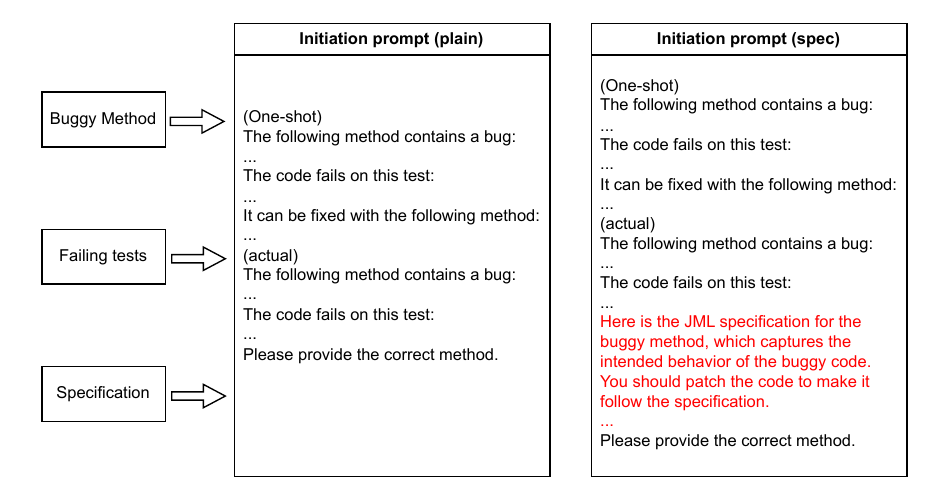}
   \caption{Patch generation prompt, with and without specifications.}
   \label{fig:cigar-prompts}
\end{figure}

\section{Evaluation}\label{sec:eval}

We conducted experiments with our implemented prototype on the Defects4J v2.1 dataset.
The Defects4J dataset~\cite{just2014defects4j} is widely used for evaluating program repair techniques for Java. We considered all bugs in the Defects4J v2.1 dataset which can be fixed within a single method,
which yielded the dataset consisting of 257 bugs. 
As we did not have access to all the tools and latest frontier LLM models internally, we compare our approach with the original CigaR~\cite{Cigar} and use
the Llama-3.1 405B model~\cite{llama} as the backend LLM for both tools (with temp = 0). 
Note that the Defects4J dataset does not require bug localisation or test cases to reproduce the bugs. Therefore, we do not report results for those phases of our approach.
While data-leakage is an issue, \cite{LessLeak25} shows Defects4J v2.1
has a leaked ratio of 0.41\%.

\begin{table}[htb]
   \caption{Number of bugs resolved by CigaR plain-mode (\cite{Cigar}) and CigaR mixed-mode (Ours)}
   \label{tab:cigar}
   \begin{tabular}{|l|c|c|c|}
   \hline
   \hline
   Bug Type & \# of bugs & CigaR~\cite{Cigar} & Ours\\
   \hline
   Logic error & 145 & 30 & 34 \\
   Incorrect handling of edge cases & 33 & 10 & 12 \\
   Null pointer & 14 & 7 & 7\\
   Index out of bound  & 15 &  6 & 6 \\ 
   Exception/Warning not thrown & 15 & 3 & 3 \\
   Type error & 6 & 3 & 4 \\
   String manipulation error & 12 & 1 & 6 \\
   Infinite loop / Stack overflow & 5 & 2  & 1 \\
   New line-related error & 5 & 1 & 1 \\
   Integer overflow & 2 & 0 & 1 \\
   EOF related error & 1 & 0 & 0 \\
   Subclassing error & 3 & 0 & 0 \\
   Wrong exception thrown & 1 & 0 & 0 \\
   \hline
   \hline
   Total & 257 &  63  &  75 \\
   \hline
   \hline
   \end{tabular}
\end{table}

The results are presented in Table~\ref{tab:cigar}.
We categorised bugs into 13 distinct bug types (e.g., null pointer, logic error, etc.) to facilitate analysis of the impact of specification generation on each bug category.
We mark the plausible patches generated by our approach as fixing the defect if
the patch passes all the tests associated with the benchmark. Experimental results show that while CigaR can find plausible patches for 24.5\% of the bugs in the dataset, our approach can find plausible patches for 29.2\% of the bugs.
It should be noted that these results could be further improved by using state-of-the-art frontier LLM models, such as Claude 4~\cite{claude4}, Gemini 2.5 Pro~\cite{gemini25pro}, and GPT-4o~\cite{gpt4o}. However, our primary objective is to investigate the impact of formal specification on automated program repair.

The main conclusion we can draw is that specifications are not useful when the
semantics of the error is clear. For example, null pointer exception, index
out of bounds or integer overflow are well understood and do not need formal
specifications.

This is best illustrated by the fix for Cli-5. The bug report (Listing~\ref{lst:Cli-5:bugreport}) and the failing
tests make it clear that the a null pointer exception is thrown.
Our approach generates the necessary patch of generating the check for the null.
The JML specification generated (Listing~\ref{lst:Cli-5:spec}) also captures this, along with a specification
for the other behaviours (where a sub-string is returned).
Hence the specification generated for the method is not useful.

\begin{lstlisting}[caption={Bug Report: Cli-5}, label={lst:Cli-5:bugreport}]
NullPointerException in Util.stripLeadingHyphens when passed a null argument
If you try to do a hasOption(null), you get a NPE:
java.lang.NullPointerException
   at org.apache.commons.cli.Util.stripLeadingHyphens(Util.java:39)
   ...
\end{lstlisting}

\begin{lstlisting}[caption={Generated JML specification: Cli-5}, label={lst:Cli-5:spec}]
 /*@requires str != null;
   @ensures str.startsWith("--") ==> \result == str.substring(2, str.length());
   @ensures str.startsWith("-") ==> \result == str.substring(1, str.length());
   @ensures !str.startsWith("-") ==> \result == str;
   @assigns \nothing; */
\end{lstlisting}

However, generated specifications can be useful for logic errors, incorrect
handling of edge cases and String manipulation errors. For example, the JML specification generated for JacksonDatabind-99 demonstrates that the specification (Listing~\ref{lst:JacksonDatabind-99:spec}) can be used to precisely capture the String manipulation issue described in the bug report (Listing~\ref{lst:JacksonDatabind-99:bugreport}).

\begin{lstlisting}[caption={Bug Report: JacksonDatabind-99}, label={lst:JacksonDatabind-99:bugreport}]
...
new ReferenceType(new TypeFactory(new LRUMap<Object, JavaType>(0, 10000)).constructType(Object.class), new PlaceholderForType(0)).toCanonical()
yields:
   java.lang.Object<$1
while the expected value is:
   java.lang.Object<$1>
\end{lstlisting}

\begin{lstlisting}[caption={Generated JML Specification: JacksonDatabind-99}, label={lst:JacksonDatabind-99:spec}]
 /*@requires _class != null;
   @required _referencedType != null;
   @ensures \result != null;
   @ensures \result.startsWith(_class.getName() + "<");
   @ensures \result.endsWith(">");
   @assigns \nothing; */
\end{lstlisting}

In our internal experiments we have discovered that
specifications are useful for unit-test generation especially ones that test
edge cases. For example, for a binary-search tree implementation, the available tests were
very limited. Generating a JML specification enabled our approach to generate
tests that checked for various edge case such
as inserting a \lstinline|null| key, a duplicate key,
deletion from an empty tree including sequence of deletes from a single node
tree.

\subsection{Limitations of the benchmark dataset}
We show a few examples to illustrate why promising
results on benchmarks
do not necessarily translate to adoption for our internal code-bases.
The main thrust of our argument is that even if the patch passes the tests, it
may not be the correct patch because, often there are only limited number of
test cases.

The first example we consider is Codec-10. The bug report notes that instead of changing the characters at the end from
\lstinline|mb| to \lstinline|m2| it makes changes at the start.
The ground-truth patch, where it changes the regular expression
\lstinline|^mb| with \lstinline|mb$|, is correct.
However, the generated patch and the JML specification are incorrect (Listing~\ref{lst:Codec-10:patch}). But there are no test cases to distinguish the generated with the correct patch. Hence if the input is \lstinline|mb123|, the behaviour of the generated patch will be incorrect. Note that the test input \lstinline|mbmb| is not sufficient
as the first branch of the if-then-else is executed.

\begin{lstlisting}[caption={Patch with JML Specification: Codec-10}, label={lst:Codec-10:patch}]
/* @requires txt !=null;
   @ensures \result != null
   @ensures \result.startsWith("M") ==> txt.startsWith("mb");
   @ensures \result.startsWith("MP") ==> txt.startsWith("mbmb"); 
*/
if (txt.endsWith("mb")) {
    txt = txt.substring(0, txt.length() - 2) + "m2";
} else {
    txt = txt.replaceAll("^mb", "m2");
}
\end{lstlisting}

The second example we present is Codec-17 where a null pointer exception was
thrown. This was because instead of calling a local method called
\lstinline|newString|, the code invoked the constructor \lstinline|new String|.
The generated patch along with the JML specification is shown in Listing~\ref{lst:Codec-17:patch}. All the test cases check only for \lstinline|null| and hence the generated patch
passes them. However, it is incorrect from a functionality perspective.

\begin{lstlisting}[caption={Patch with JML Specification: Codec-17}, label={lst:Codec-17:patch}]
/* @requires bytes != null;
   @ensures \result != null;
   @signals (NullPointerException e) bytes == null; 
*/
public static String newStringIso8859_1(final byte[] bytes) {
  return bytes == null ? null : new String(bytes, Charsets.ISO_8859_1);
}
\end{lstlisting}

The final example we consider illustrates a mixture of deficiencies in the benchmark Cli-3. The defect is how the \lstinline|Number| object is created from a string. The ground\-truth patch
creates values which are either \lstinline|Double| or \lstinline|Long|
depending on whether the string has a decimal point.
But the relevant test cases only checks for \lstinline|Double| (i.e., \lstinline|assertEquals("number flag n", Double.valueOf(4.5), line.getOptionObject("n"));|). Hence the plausible patch generated which either returns a \lstinline|Double| or
throws a \lstinline|NumberFormatException| is accepted by the test cases.
The incorrect JML specification generated is
\lstinline|@ensures \result.doubleValue() = Double.parseDouble(str)|. 

These plausible patches give the false impression that the bugs have been fully
fixed, when in fact they have not been fixed or only partially fixed. Our
initial assumption was that failing tests would provide sufficient information
about the intended behaviour of the code. However, when failing tests are incomplete or misleading, it becomes difficult to infer the intended behaviour.  Hence, a human in the loop is necessary who can correct either the generated patch or generated specification.

\section{Conclusion and Future Work}

In this paper we have presented our approach to automatic program repair. The
process we have, uses
a number of tools to mimic the standard software engineering process.
We have evaluated our approach on well-known benchmarks and highlighted the
limitations in translating the results to real codebases.
The main conclusion of our study is that techniques that work well on
standard benchmarks do not necessarily perform well on real systems without a
human in the loop.

We outline some of the ideas that we are currently exploring.
Our current use of agents is simplistic and we are investigating
the use of contract automata \cite{ContractAutomata24} as part of the agent
framework.
With a view to understanding the correct intent of the system with the defect,
we are also investigating the possibility of
combining programming by example and data-flow analysis of program fragments
\cite{CompositionalSynthesis25}.

In the context of the specification generation and validation, we have
situations where JML is not sufficient. More complex predicates and hence
verification tools like
Java Path Finder \cite{JavaPathFinder} are necessary. We are also exploring the
idea of extending the specifications with internal predicates
\cite{PredicateFix25} to narrow down the behaviour of the code
beyond method level.
We are also applying our ideas to test-case repair based on the intent (informal
specification) that we are generating from the bug report and various test cases.

Finally, since we have human in the loop, the plausible patches we generate have
to help the developer fix the defect. Hence we need to conduct a study to
understand productivity gains.

\bibliography{cutRefs}
\bibliographystyle{plain}

\end{document}